\font\twelverm=cmr10 scaled 1200    \font\twelvei=cmmi10 scaled 1200
\font\twelvesy=cmsy10 scaled 1200   \font\twelveex=cmex10 scaled 1200
\font\twelvebf=cmbx10 scaled 1200   \font\twelvesl=cmsl10 scaled 1200
\font\twelvett=cmtt10 scaled 1200   \font\twelveit=cmti10 scaled 1200
\font\twelvesc=cmcsc10 scaled 1200  
\skewchar\twelvei='177   \skewchar\twelvesy='60
     
     
\def\twelvepoint{\normalbaselineskip=12.4pt plus 0.1pt minus 0.1pt
  \abovedisplayskip 12.4pt plus 3pt minus 9pt
  \belowdisplayskip 12.4pt plus 3pt minus 9pt
  \abovedisplayshortskip 0pt plus 3pt
  \belowdisplayshortskip 7.2pt plus 3pt minus 4pt
  \smallskipamount=3.6pt plus1.2pt minus1.2pt
  \medskipamount=7.2pt plus2.4pt minus2.4pt
  \bigskipamount=14.4pt plus4.8pt minus4.8pt
  \def\rm{\fam0\twelverm}          \def\it{\fam\itfam\twelveit}%
  \def\sl{\fam\slfam\twelvesl}     \def\bf{\fam\bffam\twelvebf}%
  \def\mit{\fam 1}                 \def\cal{\fam 2}%
  \def\sc{\twelvesc}               \def\tt{\twelvett}
  \def\sf{\twelvesf}
  \textfont0=\twelverm   \scriptfont0=\tenrm   \scriptscriptfont0=\sevenrm
  \textfont1=\twelvei    \scriptfont1=\teni    \scriptscriptfont1=\seveni
  \textfont2=\twelvesy   \scriptfont2=\tensy   \scriptscriptfont2=\sevensy
  \textfont3=\twelveex   \scriptfont3=\twelveex  \scriptscriptfont3=\twelveex
  \textfont\itfam=\twelveit
  \textfont\slfam=\twelvesl
  \textfont\bffam=\twelvebf \scriptfont\bffam=\tenbf
  \scriptscriptfont\bffam=\sevenbf
  \normalbaselines\rm}
     

     
\def\beginlinemode{\endmode
  \begingroup\parskip=0pt \obeylines\def\\{\par}\def\endmode{\par\endgroup}}
\def\beginparmode{\endmode
  \begingroup \def\endmode{\par\endgroup}}
\let\endmode=\par
{\obeylines\gdef\
{}}
\def\singlespace{\baselineskip=\normalbaselineskip}

\def\oneandahalfspace{\baselineskip=\normalbaselineskip
  \multiply\baselineskip by 3 \divide\baselineskip by 2}
\def\doublespace{\baselineskip=\normalbaselineskip \multiply\baselineskip by 2}

\newcount\firstpageno
\firstpageno=2
\footline={\ifnum\pageno<\firstpageno{\hfil}\else{\hfil\twelverm\folio\hfil}\fi}
\def\toppageno{\global\footline={\hfil}\global\headline
  ={\ifnum\pageno<\firstpageno{\hfil}\else{\hfil\twelverm\folio\hfil}\fi}}
\let\rawfootnote=\footnote              
\def\footnote#1#2{{\rm\singlespace\parindent=0pt\parskip=0pt
  \rawfootnote{#1}{#2\hfill\vrule height 0pt depth 6pt width 0pt}}}
\def\raggedcenter{\leftskip=4em plus 12em \rightskip=\leftskip
  \parindent=0pt \parfillskip=0pt \spaceskip=.3333em \xspaceskip=.5em
  \pretolerance=9999 \tolerance=9999
  \hyphenpenalty=9999 \exhyphenpenalty=9999 }
\def\dateline{\rightline{\ifcase\month\or
  January\or February\or March\or April\or May\or June\or
  July\or August\or September\or October\or November\or December\fi
  \space\number\year}}
\def\received{\vskip 3pt plus 0.2fill
 \centerline{\sl (Received\space\ifcase\month\or
  January\or February\or March\or April\or May\or June\or
  July\or August\or September\or October\or November\or December\fi
  \qquad, \number\year)}}
     
     
\hsize=6.5truein
\vsize=8.5truein  
\parskip=\medskipamount
\def\\{\cr}
\twelvepoint            
\doublespace            
\overfullrule=0pt       

\def\title                      
  {\null\vskip 3pt plus 0.2fill
   \beginlinemode \doublespace \raggedcenter \bf}
     
\def\author                     
  {\vskip 3pt plus 0.2fill \beginlinemode
   \singlespace \raggedcenter\sc}
     
\def\affil                      
  {\vskip 3pt plus 0.1fill \beginlinemode
   \oneandahalfspace \raggedcenter \sl}
     
\def\abstract                   
  {\vskip 3pt plus 0.3fill \beginparmode
   \singlespace ABSTRACT: }
     
\def\endtopmatter               
  {\endpage                     
   \body}
     
\def\body                       
  {\beginparmode}               
     
\def\head#1{                    
  \goodbreak\vskip 0.5truein    
  {\immediate\write16{#1}
   \raggedcenter \uppercase{#1}\par}
   \nobreak\vskip 0.25truein\nobreak}

\def\beginitems{
\par\medskip\bgroup\def\i##1 {\item{##1}}\def\ii##1 {\itemitem{##1}}
\leftskip=36pt\parskip=0pt}
\def\enditems{\par\egroup}
     
\def\beneathrel#1\under#2{\mathrel{\mathop{#2}\limits_{#1}}}
     
\def\refto#1{$^{#1}$}           
     
\def\references                 
  {\head{References}            
   \beginparmode
   \frenchspacing \parindent=0pt \leftskip=1truecm
   \parskip=8pt plus 3pt \everypar{\hangindent=\parindent}}

\gdef\refis#1{\item{#1.\ }}                     
     
\gdef\journal#1, #2, #3, 1#4#5#6{               
    {\sl #1~}{\bf #2}, #3 (1#4#5#6)}            

\gdef\refa#1, #2, #3, #4, 1#5#6#7.{\noindent#1, #2 {\bf #3}, #4 (1#5#6#7).\rm} 

\gdef\refb#1, #2, #3, #4, 1#5#6#7.{\noindent#1 (1#5#6#7), #2 {\bf #3}, #4.\rm} 

\def\pr{\journal Phys.Rev., }

\def\jmp{\journal J.Math.Phys., }
     
\def\rmp{\journal Rev.Mod.Phys., }

\def\np{\journal Nucl.Phys., }
     
\def\pl{\journal Phys.Lett., }

\def\annp{\journal Ann.Phys.(N.Y.), }

\def\endreferences{\body}

\def\endpage                    
  {\vfill\eject}
     
\def\endpaper                   
  {\endmode\vfill\supereject}

\def\ref#1{Ref.~#1}                     
\def\Ref#1{Ref.~#1}                     
\def\[#1]{[\cite{#1}]}
\def\cite#1{{#1}}
\def\(#1){(\call{#1})}
\def\call#1{{#1}}
\def\taghead#1{}
\def\frac#1#2{{#1 \over #2}}
\def\half{{\frac 12}}

\def\12{{1\over2}}

\catcode`@=11
\newcount\r@fcount \r@fcount=0
\newcount\r@fcurr
\immediate\newwrite\reffile
\newif\ifr@ffile\r@ffilefalse
\def\w@rnwrite#1{\ifr@ffile\immediate\write\reffile{#1}\fi\message{#1}}

\def\writer@f#1>>{}
\def\referencefile{
  \r@ffiletrue\immediate\openout\reffile=\jobname.ref%
  \def\writer@f##1>>{\ifr@ffile\immediate\write\reffile%
    {\noexpand\refis{##1} = \csname r@fnum##1\endcsname = %
     \expandafter\expandafter\expandafter\strip@t\expandafter%
     \meaning\csname r@ftext\csname r@fnum##1\endcsname\endcsname}\fi}%
  \def\strip@t##1>>{}}

\def\citeall#1{\xdef#1##1{#1{\noexpand\cite{##1}}}}
\def\cite#1{\each@rg\citer@nge{#1}}	

\def\each@rg#1#2{{\let\thecsname=#1\expandafter\first@rg#2,\end,}}
\def\first@rg#1,{\thecsname{#1}\apply@rg}	
\def\apply@rg#1,{\ifx\end#1\let\next=\relax
\else,\thecsname{#1}\let\next=\apply@rg\fi\next}

\def\citer@nge#1{\citedor@nge#1-\end-}	
\def\citer@ngeat#1\end-{#1}
\def\citedor@nge#1-#2-{\ifx\end#2\r@featspace#1 
  \else\citel@@p{#1}{#2}\citer@ngeat\fi}	
\def\citel@@p#1#2{\ifnum#1>#2{\errmessage{Reference range #1-#2\space is bad.}%
    \errhelp{If you cite a series of references by the notation M-N, then M and
    N must be integers, and N must be greater than or equal to M.}}\else%
 {\count0=#1\count1=#2\advance\count1 by1\relax\expandafter\r@fcite\the\count0,
  \loop\advance\count0 by1\relax
    \ifnum\count0<\count1,\expandafter\r@fcite\the\count0,%
  \repeat}\fi}

\def\r@featspace#1#2 {\r@fcite#1#2,}	
\def\r@fcite#1,{\ifuncit@d{#1}
    \newr@f{#1}%
    \expandafter\gdef\csname r@ftext\number\r@fcount\endcsname%
                     {\message{Reference #1 to be supplied.}%
                      \writer@f#1>>#1 to be supplied.\par}%
 \fi%
 \csname r@fnum#1\endcsname}
\def\ifuncit@d#1{\expandafter\ifx\csname r@fnum#1\endcsname\relax}%
\def\newr@f#1{\global\advance\r@fcount by1%
    \expandafter\xdef\csname r@fnum#1\endcsname{\number\r@fcount}}

\let\r@fis=\refis			
\def\refis#1#2#3\par{\ifuncit@d{#1}
   \newr@f{#1}%
   \w@rnwrite{Reference #1=\number\r@fcount\space is not cited up to now.}\fi%
  \expandafter\gdef\csname r@ftext\csname r@fnum#1\endcsname\endcsname%
  {\writer@f#1>>#2#3\par}}

\def\ignoreuncited{
   \def\refis##1##2##3\par{\ifuncit@d{##1}%
    \else\expandafter\gdef\csname r@ftext\csname r@fnum##1\endcsname\endcsname%
     {\writer@f##1>>##2##3\par}\fi}}

\def\r@ferr{\endreferences\errmessage{I was expecting to see
\noexpand\endreferences before now;  I have inserted it here.}}
\let\r@ferences=\references
\def\references{\r@ferences\def\endmode{\r@ferr\par\endgroup}}

\let\endr@ferences=\endreferences
\def\endreferences{\r@fcurr=0
  {\loop\ifnum\r@fcurr<\r@fcount
    \advance\r@fcurr by 1\relax\expandafter\r@fis\expandafter{\number\r@fcurr}%
    \csname r@ftext\number\r@fcurr\endcsname%
  \repeat}\gdef\r@ferr{}\endr@ferences}


\let\r@fend=\endpaper\gdef\endpaper{\ifr@ffile
\immediate\write16{Cross References written on []\jobname.REF.}\fi\r@fend}

\catcode`@=12

\citeall\refto		
\citeall\ref		%
\citeall\Ref		%

\def\s{\sigma}
\def\half{{1 \over 2}}
\def\ra{{\rangle}}
\def\la{{\langle}}

\def\ih{{i \over \hbar}}

\def\E{{\cal E}}

\def\ria{{\rightarrow}}

\def\sp{{\sigma_{\!{\fiverm +}}}}
\def\sm{{\sigma_{\!{\fiverm -}}}}
\def\tsp{{{\tilde \sigma}_{\!{\fiverm +}}}}
\def\tsm{{{\tilde \sigma}_{\!{\fiverm -}}}}

\centerline{\bf Arrival Times in Quantum Theory}
\centerline{\bf from an Irreversible Detector Model}

\author J.J.Halliwell 
\vskip 0.2in
\affil
Theory Group
Blackett Laboratory
Imperial College
London SW7 2BZ
UK
\vskip 0.5in
\centerline {\rm Preprint IC/TP/97-98/49. June, 1999}
\vskip 0.1in
\vskip 1.0in
\centerline {\rm  PACS numbers: 03.65.-w, 03.65.Bz, 06.30.Ft}
\vskip 1.0in

\abstract {We investigate a detector scheme designed to measure the
arrival of a particle at $x=0$ during a finite time interval. The
detector consists of a two state system which  undergoes a
transition from one state to the other when the particle crosses
$x=0$, and possesses the realistic feature that it is effectively
irreversible  as a result of being coupled to a large environment. 
The probabilities for crossing or not crossing $x=0$ thereby derived
coincide with earlier phenomenologically proposed expressions
involving a complex potential. The probabilities are compared with
similar previously proposed expressions involving sums over paths,
and a connection with time operator approaches is
also indicated.}

\endtopmatter 
\endpage

\head{\bf 1. Introduction}

An enduring class of questions in non-relativistic quantum theory 
are those that involve time in a non-trivial way
[\cite{All,Time,GRT,AOP}]. Of these, the question of tunneling time is
perhaps the most important  [\cite{HaS,Lan}]. But some of the basic
issues are most simply seen through the question, what is the
probability that a particle enters a region of space for the first
time during a given time interval?  What makes this sort of problem
interesting is that standard quantum mechanics seems to handle it
with some difficulty and there does not appear to be a unique
answer -- classically equivalent methods of assigning the arrival
time can differ at the quantum level.

One of the key sources of difficulty with defining arrival times in
quantum theory is that they involve specification of positions at
two moments of time: to state that the particle enters a given
spatial region at time $t$ means that it is outside the region
immediately before $t$ {\it and} inside it immediately after $t$.
Since positions at different moments of time do not commute, we do
not expect to be able to associate a single hermitian operator  with
the arrival time. Of course, many different methods of defining
arrival time then naturally suggest themselves, but, like the
question of specifying phase space locations in quantum mechanics,
they are not all equivalent.

To be specific, in this paper we consider the following question:
Suppose we have an initial state with support only in $x>0$. Then
what is the probability that the particle enters the region $x<0$ 
at any time during the time interal $ [ 0, \tau ] $?

A number of previous papers have addressed this particular problem
using path integrals [\cite{YaT,Har,MiH}]. (These approaches are
closely related to the decoherent histories approach to quantum
theory [\cite{GeH,DH}]). The amplitudes for crossing and not
crossing $x=0$ are obtained by summing over paths which either
always enter or never enter $x<0$, and probabilities are then
obtained by squaring amplitudes in the normal way. However, due to
interference between paths, the resultant probabilities do not add
up to $1$, so cannot be regarded as true probabilities. A way past
this difficulty was explored in Ref.[\cite{HaZa}]. There, the point
particle system was coupled to a thermal environment to induce
decoherence of different paths in configuration space, and the
correct probability sum rules were restored.  Although this approach
produced mathematically viable candidates for the probabilities of
crossing and not crossing, they depend to some degree on the
environment, and it is by no means clear how the results  correspond
to a particular type of measurement, even an idealized one. General
theorems exist showing that decoherence of histories implies the
existence at the final end of the histories of a record storing the
information about the decohered histories [\cite{GeH}],  but these
have been explicitly found only in a few simple cases (see
Ref.[\cite{Hal4}], for example). For these reasons, it is of
interest to compare these path integral approaches with a completely
different approach involving a specific model of a detector.

Let us therefore introduce a model detector which is coupled to the
particle in the region $ x<0$, and such that it undergoes a
transition when the coupling is switched on.  Such an approach has
certainly been considered before (see, {\it e.g.},
Ref.[\cite{AOP}]). The particle could, for example, be coupled to a
simple two-level system that flips from one level to the other when
the particle is detected. One of the difficulites of many detector
models, however, is that if they are modeled by unitary quantum
mechanics, the possibility of the reverse transition exists. Because
quantum mechanics is fundamentally reversible, the detector could 
return to the undetected state under its self-dynamics,  even when
the particle has interacted with it. 

To get around this difficulty, we appeal to the fact that realistic
detectors have a very large number of degrees of freedom, and are
therefore effectively {\it irreversible}. They are designed so that
there is an overwhelming large probability for them to make a
transition in one direction rather than its reverse.  In this paper
we introduce a simple model detector that has this property.  This
is achieved by coupling a two-level system detector to a large
environment, which makes its evolution effectively irreversible. The
description of this system is easily obtained using some standard
machinery of open quantum systems, and the resulting master equation
for the particle coupled to the detector actually has the Lindblad
form [\cite{Lin}]. 

We are not concerned with a specific experimental arrangement, but
rather, as is common in quantum measurement theory, an idealized
model which has as many physically realistic features one can
reasonably incorporate. In particular, in contrast to most
measurement models discussed in the literature, it has the key
property of irreversibility.

The detector is described in Section II. On solving the detector
dynamics, an expression for the probability of entering a spacetime
region is obtained. It has the  appearance of a probability obtained
from a wave function satisfying a Schr\"odinger equation with an
imaginary contribution to the potential, which has previously been
proposed as a phenomenological device [\cite{All,MBM,PMB}].  Our
detector model therefore justifies previously used phenomenological
approaches. 

The probabilities obtained are also readily compared with the
results of the path integral approaches, and the comparison sheds
some light on the shortcomings of the latter. This is described in
Section III. A brief mention is also made of the possible connection
with  time operators.

\head{\bf 2. The Detector Model}

In this section we describe a detector designed to make a
permanent record of whether or not a particle enters the region
$x<0$ at any time during a given finite time interval.
We take the detector to be a two-level system, with levels $ | 1 \ra
$ and $ | 0 \ra $, representing the states of no detection and
detection, respectively. Introduce the raising and lowering
operators 
$$
\sp = | 1 \ra \la 0 |, \quad \sm = | 0 \ra \la 1 |
\eqno(2.1)
$$
and let the Hamiltonian of the detector be $H_d = \half \hbar
\omega \s_z $, where 
$$ 
\sigma_z = | 1 \ra \la 1 | - | 0 \ra \la 0 |
\eqno(2.2)
$$
so $ | 0 \ra $ and $ | 1 \ra $ are eigenstates of $H_d$ with
eigenvalues $ - \half \hbar \omega $ and $ \half \hbar \omega $
resepectively. We would like to couple the detector to a free
particle in such a way that the detector makes an essentially 
irreversible transition from $ | 1 \ra $ to $ | 0 \ra $ if the
particle enters $x < 0 $, and remains in $ | 1 \ra $ otherwise. This
can be arranged by coupling the detector to a large environment of
oscillators in their ground state, with a coupling proportional to $
\theta (-x)$. This means that if the particle enters the region
$x<0$,  the detector becomes coupled to the large environment
causing it to undergo a transition. Since the environment is in its
ground state, if the detector initial state is the higher energy
state $ | 1 \ra $ it will, with overwhelming probability, make a
transition from $ | 1\ra $ to the lower energy state $ | 0 \ra $.
A possible Hamiltonian describing this
process for the three-component system is
$$
H = H_s + H_d + H_{\E} + V(x)  H_{d \E }
\eqno(2.3)
$$
where the first three terms are the  Hamiltonians of the particle,
detector and environment respectively, and $H_{d \E}$ is the
interaction Hamiltonian of the detector and its environment.  
The simplest choice
of environment is a collection of harmonic
oscillators,
$$
H_{\E} = \sum_n \hbar \omega_n a_n^{\dag} a_n
\eqno(2.4)
$$
and we take the coupling to the detector to be via the interaction
$$
H_{d \E} = \sum_n \hbar \left( \kappa^*_n \sm a_n^{\dag} +
\kappa_n \sp a_n \right)
\eqno(2.5)
$$
An environment consisting of an electromagnetic field, for example,
would give terms of this general form. 
$V(x)$
is a potential concentrated in $x<0$ (and we will eventually make
the simplest choice, $ V(x) = \theta (-x) $, but for the moment we
keep it more general). The important feature is that the interaction
between the detector and its environment, causing the detector to
undergo a transition, is switched on only when the particle is in
$x<0$.

A similar although more
elaborate model particle detector has been previously studied by
Schulman [\cite{Sch1}] (see also Refs.[\cite{Sch2}]).  It consists
of a lattice of spins in a metastable state interacting with the
lattice's phonon modes, with interactions essentially the same as in
Eq.(2.3). His approach has the advantage that the amplification of a
microscopic event as well as the irreversibility of the measurement
is modeled, but it does not appear to be possible to solve it as
explicitly as the simpler model considered here. For an 
even more elaborate model see Ref.[\cite{Gur}].

We are interested in the reduced dynamics of the particle and
detector with the environment traced out. Hence we seek a master
equation for the reduced density operator $\rho$ of the particle and
detector. With the above choices for $H_{\E}$ and $H_{d \E} $, the
derivation of the master equation is standard [\cite{Car,Gar}] and
will not be repeated here. There is the small complication of the
factor of $V(x)$ in the interaction term, but this is readily
accommodated. We assume a factored initial state, and we assume that
the environment starts out in the ground state. In a Markovian
approximation (essentially the assumption that the environment
dynamics is much faster than detector or particle dynamics), and in
the approximation of weak detector-environment coupling, the master
equation is
$$
\dot \rho = - \ih [ H_s + H_d, \rho] 
- { \gamma \over 2} \left( V^2 (x ) \sp \sm \rho \ +  \rho
\sp \sm  V^2 (x)  \ -  \ 2 V (x) \sm \rho \sp V (x )
\right) 
\eqno(2.6)
$$
Here, $\gamma$ is a phenemonological constant determined  by the
distribution of oscillators in the environment and underlying
coupling constants. The frequency $\omega $ in $H_d$ is also
renormalized to a new value $\omega'$.

Eq.(2.6) is the sought-after description of a particle coupled to an
effectively irreversible detector in the region $ x< 0 $. In the
dynamics of the detector plus environment only ({\it i.e.}, with
$V=1$ and $H_s=0$), it is readily shown that every initial state
tends to the state $ | 0 \ra \la 0 | $ on a timescale $\gamma^{-1}$.
With the particle coupled in, as in (2.6), if the initial state of
the detector is chosen to be $ | 1 \ra \la 1 | $, it undergoes an
irreversible transition to the state $ | 0 \ra \la 0 | $ if the
particle enters $ x < 0 $, and remains in its initial state
otherwise.

Although we have outlined the derivation of this master equation for
a particular choice of environment and detector-environment
coupling, we expect that the form of the equation is more general. 
It is well-known that, after tracing out the environment, and in the
approximation that the evolution is Markovian, the reduced density
operator $\rho $ of the particle and detector must evolve according
to a master equation of the Lindblad form 
$$
\dot \rho = - \ih [ H_s + H_d , \rho ] + \sum_m \left( L_m \rho
L_m^{\dag} - \half L_m^{\dag} L_m \rho - \half \rho L_m^{\dag} L_m
\right) 
\eqno(2.7)
$$
This is the most general Markovian evolution equation preserving the
positivity, hermiticity and trace of $\rho $. The operators $L_m$
model the effects of the environment [\cite{Lin}].  The form (2.6)
is recovered with a single Lindblad operator $L = \gamma^{\half}
V(x) \sm $. A similar detection scheme based on a postulated master
equation similar to (2.7) was previously  considered in
Ref.[\cite{Jad,BlJ}], although the resultant expressions for arrival
time probability given below were not derived (and no microscopic
origin of the equation was given).
  
Eq.(2.6) is easily solved by writing
$$
\rho = 
\rho_{11} \otimes | 1 \ra \la 1 |
+ \rho_{01} \otimes | 0 \ra \la 1 |
+ \rho_{10} \otimes | 1 \ra \la 0 |
+ \rho_{00} \otimes | 0 \ra \la 0 |
\eqno(2.8)
$$
where
$$
\eqalignno{
\dot \rho_{11} &= - \ih [ H_s, \rho_{11} ]
- {\gamma \over 2} \left( V (x) \rho_{11} + \rho_{11} V (x) \right)
&(2.9) \cr
\dot \rho_{01} &= - \ih [ H_s, \rho_{01} ]
- {\gamma \over 2} \rho_{01} V (x) + i { \hbar \omega' \over 2} \rho_{01}
&(2.10) \cr
\dot \rho_{10} &= - \ih [ H_s, \rho_{10} ]
- {\gamma \over 2} V (x) \rho_{10} - i { \hbar \omega' \over 2} \rho_{10}
&(2.11) \cr
\dot \rho_{00} &= - \ih [ H_s, \rho_{00} ]
+ \gamma V (x) \rho_{11} V (x)
&(2.12) \cr }
$$
(where we have now set $V(x) = \theta (-x)$, so that $V^2 = V$).
We suppose that the particle starts out in an initial state $ |
\Psi_0 \ra $, hence the above equations are to be solved subject to
the initial condition,
$$
\rho ( 0 ) = | \Psi_0 \ra \la \Psi_0 | \otimes | 1 \ra \la 1 |
\eqno(2.13)
$$

The probability of finding the detector in the
unregistered state $ | 1 \ra $ at time $\tau $ is
$$
p_{nd} = {\rm Tr} \rho_{11} (\tau ) = \int_{- \infty}^{\infty} dx 
\ \rho_{11}(x,x, \tau )
\eqno(2.14)
$$
and the probability of finding it in the registered state
$ | 0 \ra $, is
$$
p_d = {\rm Tr} \rho_{00} ( \tau ) = \int_{- \infty}^{\infty} dx 
\ \rho_{00}(x,x, \tau )
\eqno(2.15)
$$
(where the trace is over the particle Hilbert space).
Clearly $p_{nd} + p_d = 1 $, since $ {\rm Tr} \rho = 1 $.

Note that the probability for no detection includes an integral over
$ x<0 $ and $\rho_{11} (x,x, \tau ) $ is not necessarily zero for
$x<0$. There is therefore the possibility  that the particle could
enter the region $x<0$ without the  detector registering the fact.
This is however, to be expected of a realistic detector -- there is
some probability that it will fail to do what it is supposed to do.
The probability of this happening is typically small, and indeed,
computation of this probability provides a useful check on the
efficiency of the detector (although below we will check detector
efficiency in a different way).

The formal solution to Eq.(2.9) for $\rho_{11}$ may be written
$$
\rho_{11} (t) = \exp \left( - \ih H_s t - { \gamma \over 2} V t \right) 
\ \rho_{11} (0)
\ \exp \left( \ih H_s t - { \gamma \over 2} V  t \right)
\eqno(2.16)
$$
What is particularly interesting about this expression is that it
can be factored into a pure state. Let 
$ \rho_{11} = | \Psi \ra \la \Psi | $. Then, noting that
$ \rho_{11} (0) = | \Psi_0 \ra \la \Psi_0 | $,
Eq.(2.16) is equivalent to
$$
| \Psi (t) \ra = \exp \left( - \ih H_s t - { \gamma \over 2} V t \right)
| \Psi_0 \ra
\eqno(2.17)
$$
The probability for no detection is then
$$
p_{nd} 
= \int_{-\infty}^{\infty} dx \ | \Psi (x, \tau ) |^2
\eqno(2.18)
$$
The pure state (2.17) evolves according to a Schr\"odinger equation
with an imaginary contribution to the potential, $ - \half i \hbar
\gamma V $. Complex potentials have been used previously in this
context, as phenomenological devices, to imitate absorbing boundary
conditions (see, for example Refs.[\cite{All,MBM,PMB}]).
Here, the appearance of a complex potential is {\it derived}
from the master equation of a particle coupled to an irreversible
detector, which in turn may be derived from the unitary dynamics of
the combined particle--detector--environment system.

Eq.(2.12) for $\rho_{00} $ may be formally solved to yield
$$
\rho_{00} (t) = \gamma \int_0^t dt' \exp \left( - \ih H_s (t-t') \right)
V (x) \rho_{11} (t') V (x)
\exp \left(  \ih H_s (t-t') \right)
\eqno(2.19)
$$
(recalling that $\rho_{00} (0) = 0 $). Inserting the solution
for $\rho_{11} (t') $, the probability for detection may be written,
$$
\eqalignno{
p_d &= 
\gamma \int_0^\tau dt \int_{- \infty}^{\infty} dx \left|
\la x | \exp \left( - \ih H_s (\tau -t) \right) V(x) 
\exp \left( - \ih H_s t - { \gamma \over 2} V t \right) | \Psi_0 \ra
\right|^2
\cr
&=
\gamma \int_0^\tau dt \int_{- \infty}^0 dx \left|
\la x | 
\exp \left( - \ih H_s t - { \gamma \over 2} V t \right) | \Psi_0 \ra
\right|^2
\cr
&=
\gamma \int_0^\tau dt \int_{- \infty}^0 dx \left| 
\Psi (x, t ) \right|^2
&(2.20) \cr}
$$
where $\Psi (x,t) $ is the wave function (2.17). The expression for the
probability for detection has an appealing form: it is the integral
of $ | \Psi (x,t) |^2 $ over the space-time region of interest. 
It is crucially important, however, that the wave function satisfies
not the usual Schr\"odinger equation, but one with an imaginary
contribution to the potential modeling the detector. 

It is useful to write the probabilities for detection and
no detection in the form,
$$
p_d = {\rm Tr} \left( \Omega \ | \Psi_0 \ra \la \Psi_0 | \right),
\quad p_{nd} = {\rm Tr} \left( \bar \Omega 
\ | \Psi_0 \ra \la \Psi_0 | \right)
\eqno(2.21)
$$
where
$$
\eqalignno{
\Omega &= \int_0^{\tau} dt
\ \exp \left( \ih H_s t  - { \gamma \over 2} V t \right)
\ \gamma V
\ \exp \left( -\ih H_s t - { \gamma \over 2} V t \right)
&(2.22) \cr
\bar \Omega &= 
\ \exp \left( \ih H_s \tau - { \gamma \over 2} V \tau \right)
\ \exp \left( - \ih H_s \tau - { \gamma \over 2} V \tau \right)
&(2.23) \cr}
$$
The first of these expressions follows from (2.20)
from the properties of the trace, and using the fact that
$ V^2 = V $.

$\Omega $ and $\bar \Omega $ are clearly not projection operators,
although their properties are close to those of projectors. They are
both positive operators and $\Omega + \bar \Omega = 1$. The latter
follows by integrating the identity
$$
{ d \bar \Omega \over d \tau } =
- \ \exp \left( \ih H_s \tau - { \gamma \over 2} V \tau \right)
\ \gamma V \ \exp \left( -\ih H_s \tau - { \gamma \over 2} V \tau \right)
\eqno(2.24)
$$
Moreover, these operators clearly have the desired localization
properties on histories of particle positions, as will be seen most
clearly in the path integral expression of the next section. We do
not expect to be able to associate a true projection operator with
the arrival time, but here we have found a POVM, which is the next
best thing. 

Eqs.(2.21)--(2.23) are the main result of this section: expressions
for the probabilities of entering or not entering a region
of spacetime, derived from an irreversible detector model.

We now consider the issue of the efficiency of the detector. A
simple way to  do this is to introduce a second detector identical
to the first and located in the region $x>0$. Since the entire $x$-axis
is now monitored, the probability that neither detector registers
during the time interval is then a measure of the degree to which
the detector will fail.

With two detectors in place, the master equation now is,
$$
\eqalignno{
\dot \rho = - \ih [ H_s , \rho] 
&- { \gamma \over 2} \left( \theta (-x) \sp \sm  \rho \ +  \ \rho
\sp \sm \theta (-x)   \ -  \ 2 \theta (-x) \sm \rho \sp \theta (-x)
\right) 
\cr
&- { \gamma \over 2} \left( \theta (x) \tsp \tsm 
\rho \ + \ \rho
\tsp \tsm \theta (x)   
\ - \ 2 \theta (x) \tsm \rho \tsp \theta (x)
\right) 
&(2.25) \cr}
$$
where $ \tsp, \tsm $ 
are the raising and lowering operators for the
detector in $x>0$. This equation is solved like (2.6),
by writing
$$
\rho = \rho_{nd} \otimes | 1 \ra \la 1 | \otimes | 1 \ra \la 1 |
+ \cdots
\eqno(2.26)
$$
We are interested only in the probability that neither detector
registers, so we omit the explicit form of the other terms in (2.26).
It is readily shown that
$$
\dot \rho_{nd} = - \ih [ H_s, \rho_{nd} ] - \gamma \rho_{nd}
\eqno(2.27)
$$
The probability of no detection is $ {\rm Tr} \rho_{nd} $, and from
(2.27), it clearly decays like $e^{ - \gamma t } $. Hence the
detector functions efficiently if the total time duration $\tau$ is
much greater than $\gamma^{-1}$. (The potential inefficiency of the
detector for $\tau$ not sufficiently large compared to $\gamma^{-1}$
corresponds to the difficulties in
defining the time operator at low momenta [\cite{GRT}].)

The evolution according to (2.17) for the case
in which $V(x)$ is a real step function was studied in detail by
Allcock [\cite{All}], who was consequently  pessimistic about the
possibility of defining arrival time.  This is partly because
$\gamma $ needs to be large for accurate detection, but in this case
reflection from the potential is high and not all of the incoming
flux is absorbed. Subsequent authors have shown that his pessimism
was misplaced, if more general potentials $\tilde V(x)$ are permitted,
which are smoother and may also need to be complex (see, for
example, Refs.[\cite{MBM,PMS}]). The considerations of this paper
readily generalize to this case by replacing Eq.(2.6) with the
equation
$$
\dot \rho = - \ih [ H_s , \rho] 
- { \gamma \over 2} \left( \tilde V^{\dag} \tilde V
\sp \sm \rho \ + \ \rho
\sp \sm \tilde V^{\dag} \tilde V  \ - 
\ 2 \tilde V  \sm \rho \sp \tilde V^{\dag}
\right) 
\eqno(2.28)
$$
which is still of the Lindblad form, but $\tilde V (x)$ is generally a
non-hermitian operator.  It is less clear what sort of
detector-environment coupling this corresponds to in Eq.(2.3) and
this is certainly of interest to investigate. Note also that $\tilde V^2
\ne \tilde V $, but the above results are readily modified to incorporate
this.

The detector described here measures whether the particle  entered
the region $x<0$ at any time during $ [0, \tau]$, for $\tau >>
\gamma^{-1}$. It could easily be extended to give more precise
information about the time at which the particle enters $x<0$ by
using a series of similar detectors, but with a time-dependent
coupling to the environment, so that the detectors can be switched
on or off  at a succession of times $t_1, t_2, t_3 \cdots $
(separated in time by at least $\gamma^{-1}$) in the interval $ [ 0,
\tau ]$.

\head{\bf 3. Comparison with Other Approaches}

The expressions we have derived for detection and no detection bear
a close resemblance to  previously derived path integral expressions
for the probabilities of entering or not entering the region
$x<0$, so we now carry out a comparison.

For simplicity let the initial and final points lie in $x>0$.
The amplitude for remaining restricted to the region $x>0$ is
$$
g_r (x_f ,\tau | x_0, 0 ) = \int_r {\cal D} x (t)
\exp \left( \ih S [ x (t) ]  \right)
\eqno(3.1)
$$
where the sum is over all paths in $ x> 0 $. The amplitude
for crossing $x=0$ at some time in the interval $ [ 0, \tau ]$ is
$$
g_c (x_f ,\tau | x_0, 0 ) = \int_c {\cal D} x (t)
\exp \left( \ih S [ x (t) ]  \right)
\eqno(3.2)
$$ 
where the sum is over paths that spend some time in $x<0$
(see Refs.[\cite{PDX,Hal3,Hal5}] for details of construction
of these objects).  
Clearly $ g_r + g_c = g $, where $g$ denotes the unrestricted
propagator, given by a sum over all paths.
The probabilities for crossing and not crossing $x=0$ are then
obtained from these propagators, by attaching an initial state,
squaring the amplitudes, and then summing over final positions in
the usual way. However, as stated earlier, the resultant probabilities for
crossing and not crossing computed in this way do not sum to $1$, 
because of interference between the different types of paths.

Now we compare with the measurement model of the previous section.
The probabilities computed here for detection or no detection in the
region $x<0$ automatically sum to $1$. The probability for no
detection may computed from (2.17).  The evolution operator that
appears there may be written in path integral form, 
$$ \eqalignno{
g_{nd} (x_f, \tau | x_0, 0 ) &= \la x_f | \exp \left( - \ih H_s \tau
 - { \gamma \over 2} V \tau \right) | x_ 0 \ra \cr & = \int {\cal D}
x (t) \exp \left( \ih S [ x (t) ] -  {\gamma \over 2} \int_0^{\tau}
dt \ V ( x (t) ) \right) 
&(3.3) \cr } 
$$ 
The sum is over all paths
$x(t)$ connecting $x_0$ at time $0$ to $x_f$ at time $\tau$. But it
is clear that the potential $ V(x)$ suppresses contributions from
paths that enter $ x< 0 $. Split the paths summed over into the two 
classes $r$ and $c$, as above. (For simplicity, we take $x_0>0,
x_f>0 $). Noting that $V=0$ in $x>0$, the path integral becomes, 
$$
\eqalignno{ g_{nd} (x_f, \tau | x_0, 0 )  & = \int_r {\cal D} x (t)
\exp \left( \ih S [ x (t) ]  \right) \cr & + \int_c {\cal D} x (t)
\exp \left( \ih S [ x (t) ] -  {\gamma \over 2} \int_0^{\tau} dt \ V
( x (t) ) \right) 
&(3.4) \cr } 
$$ 
Comparing with Eq.(3.1), we see that (3.4) differs from it by the
presence of the second term. In the second term, every path in the
sum has a section lying in the region $x<0$ and an exponential
suppression factor will come into play.  $g_{nd}$ and $g_r$ exactly
coincide in the limit $\gamma \ria \infty$. The resultant 
probabilities are,
however, not very interesting [\cite{YaT}]. Furthermore, as stated,
large $\gamma$ means that most of the incoming wave packets are
reflected rather than absorbed by the detector. This means that the
second term in Eq.(3.4) will generally be significant and $g_{nd}$
and $g_r$ will not be close. From this we conclude that the purely
path integral approaches to defining the arrival time, as expressed
by (3.1) and (3.2), are actually rather removed from the more
physically motivated expressions using a model detector derived in
this paper. Including a physical mechanism for decoherence in the
path integral approach [\cite{HaZa}] yields more sensible results,
but they are not very closely related to the detector results (a
comparison is carried out in an earlier version of this paper
[\cite{Hal5}]).

We may also write the POVM for no detection, (2.23), in another
more enlightening form. Note that
$$
\bar \Omega = U^{\dag} (\tau) U (\tau)
\eqno(3.5)
$$
where
$$
\eqalignno{
U ( \tau ) &= \ \exp \left( \ih H_s \tau  \right)
\ \exp \left( - \ih H_s \tau - { \gamma \over 2} V \tau \right)
\cr 
& = T \exp \left( - { \gamma \over 2} \int_0^{\tau} dt \ V(x_t )
\right)
&(3.6) \cr }
$$
Here, $T$ denotes time ordering, and $x_t = x + pt / m $ is the
position operator at time $t$ in the Heisenberg picture.
Splitting the time interval into small units $\delta t$,
$U$ is therefore a time-ordered product of operators
of the form $ \exp ( - \gamma \delta t V / 2 ) $.
But with the choice used here, $ V(x) = \theta (-x) $,
we have $V^2 = V $ so
$$
\eqalignno{
\exp \left( - { \gamma \over 2 } \delta t V \right)
&= (1 - V ) + V e^{-\gamma \delta t / 2 }
\cr
&=\theta (x) + \theta (-x) e^{-\gamma \delta t / 2 }
&(3.7) \cr }
$$
This is therefore a projector onto the positive $x$-axis
plus an exponentially smaller projector onto the negative
$x$-axis. Naive expectations would lead one to guess the
first term, but the addition of the second one appears to
be important in making the probabilities well-defined
(see Ref.[\cite{Hal3}] for related formulae).

From the above we see that the probabilities for detection
and no detection depend on $x$ and $p$ only through the
operators $ \theta (x_t) $ at a series of times.
This leads us to a connection with time operators.
For classical quantities $x$, $p$, we have
$$
\theta ( x + {p t \over m} ) = \theta ( {m x \over p} + t )
\eqno(3.8)
$$
(for $x,p > 0 $).
The point is that the quantity $ m x / p $ is precisely the
classical arrival time. It is the quantity that numerous authors
have, not without serious difficulties,  attempted to elevate to the
status of a quantum operator to define quantum arrival times (see,
for example, Ref.[\cite{GRT}]). The connection between the expressions
derived here and those derived using the time operator may therefore
be obtained by investigating the extent to which the classical
relation (3.8) persists in the quantum theory. This will be
pursued elsewhere.

\head{\bf 4. Summary and Conclusions}

We have presented a detector model for the measurement of arrival
time in quantum theory. It possesses the realistic feature of being
effectively irreversible. The results of the scheme connect very
nicely with previous approaches involving a postulated complex
potential to imitate the effects of  a detector.

The detector model was compared with previous approaches involving
sums over paths, and the detector model exposes the limitations
of the latter. Some indication of the possible connection with time
operators was also given.

\head{\bf Acknowledgements}

This paper was inspired by an invitation to speak at the workshop,
{\it Time in Quantum Mechanics}, at La Laguna, Tenerife, May, 1998.
I would like to thank the organizers for giving me this opportunity,
and the other participants of the meeting, for their useful
comments. Special thanks go to Gonzalo Muga for his encouraging
comments and for his hospitality. I would also like to thank Jason
Twamley for his useful suggestions, and Larry Schulman
for helpful comments about detector models.

\references

\def\pr{{\sl Phys. Rev.\ }}

\def\jmp{{\sl J. Math. Phys.\ }}
\def\rmp{{\sl Rev. Mod. Phys.\ }}

\def\np{{\sl Nucl. Phys.\ }}
\def\pl{{\sl Phys. Lett.\ }}
\def\annp{{\sl Ann. Phys. (N.Y.)\ }}

\refis{AOP} Y.Aharanov, J.Oppenheim, S.Popescu, B.Reznik and
W.Unruh, quant-ph/9709031 (1997).

\refis{All} G.R.Allcock, \annp {\bf 53}, 253 (1969); 
{\bf 53}, 286 (1969); {\bf 53}, 311 (1969).

\refis{BlJ} Ph.Blanchard and A.Jadczyk, 
{\sl Helv.Phys.Acta.} {\bf 69}, 613 (1996).


\refis{Car} H.Carmichael, {\it An Open Systems Approach to Quantum
Optics} (Springer-Verlag Lecture Notes in Physics, m18, Berlin,
1993).


\refis{GeH} M.Gell-Mann and J.B.Hartle, {\sl Phys.Rev.} 
{\bf D47}, 3345 (1993).

\refis{Gar} C.W.Gardiner, {\it Quantum Noise} (Springer-Verlag,
Berlin, 1991).

\refis{DH} R.B.Griffiths, {\sl J.Stat.Phys.} {\bf 36}, 219 (1984);
{\sl Phys.Rev.Lett.} {\bf 70}, 2201 (1993);
{\sl Am.J.Phys.} {\bf 55}, 11 (1987);
R.Omn\`es, {\sl J.Stat.Phys.} {\bf 53}, 893 (1988);
{\bf 53}, 933 (1988);
{\bf 53}, 957 (1988);
{\bf 57}, 357 (1989);
{\bf 62}, 841 (1991);
{\sl Ann.Phys.} {\bf 201}, 354 (1990); 
{\sl Rev.Mod.Phys.} {\bf 64}, 339 (1992);
J.B.Hartle, in {\it Quantum Cosmology and Baby
Universes}, S. Coleman, J. Hartle, T. Piran and S. Weinberg (eds.)
(World Scientific, Singapore, 1991);
J.J.Halliwell,
in {\it Fundamental Problems in Quantum Theory}, 
edited by D.Greenberger and A.Zeilinger,
Annals of the New York Academy of Sciences, Vol 775, 726 (1994).
For further developments in the decoherent histories approach,
particularly adpated to the problem of spacetime
coarse grainings, see
C. Isham, \jmp {\bf 23}, 2157 (1994);
C. Isham and N. Linden, \jmp {\bf 35}, 5452 (1994);
{\bf 36}, 5392 (1995);
C. Isham, N. Linden and S.Schreckenberg,
\jmp {\bf 35}, 6360 (1994).

\refis{GRT} N.Grot, C.Rovelli and R.S.Tate, \pr {\bf A54}, 4676 (1996).

\refis{Gur} S.A.Gurvitz, ``Dephasing and Collapse in Continuous
Measurement of a Single System'', quant-ph/9808058.

\refis{Hal3} J.J.Halliwell, {\sl Phys.Lett} {\bf A207}, 237 (1995).

\refis{Hal4} J.J.Halliwell, ``Somewhere in the Universe: Where is
the Information Stored when Histories Decohere?'', quant-ph/9902008,
Imperial preprint TP/98-99/29 (1999).

\refis{Hal5} J.J.Halliwell, quant-ph/9805057.

\refis{HaZa} J.J.Halliwell and E.Zafiris, 
{\sl Phys.Rev.} {\bf D57}, 3351-3364 (1998). 


\refis{Har} J.B.Hartle, {\sl Phys.Rev.} {\bf D44}, 3173 (1991).

\refis{HaS}  E.H.Hauge and J.A.Stovneng, \rmp {\bf 61}, 917
(1989).

\refis{Jad} A.Jadcyk, {\sl Prog.Theor.Phys.} {\bf 93}, 631 (1995).

\refis{Lan} R.Landauer, {\sl Rev.Mod.Phys.} {\bf 66}, 217 (1994);
{\sl Ber.Bunsenges.Phys.Chem} {\bf 95}, 404 (1991).

\refis{Lin} G.Lindblad, {\sl Commun.Math.Phys.} {\bf 48}, 119
(1976).

\refis{MiH} R.J.Micanek and J.B.Hartle, {\sl Phys.Rev.} {\bf A54},
3795 (1996). 

\refis{MBM} J.G.Muga, S.Brouard and D.Mac\'ias, \annp {\bf 240},
351 (1995).

\refis{PMS} J.P.Palao, J.G.Muga and R.Sala,
quant-ph/9805035 (1998).

\refis{PMB} J.P.Palao, J.G.Muga, S.Brouard and A.Jadczyk,
\pl {\bf A233}, 227 (1997).

\refis{PDX} A.Auerbach and S.Kivelson, \np {\bf B257}, 799 (1985);
P. van Baal,  in {\it Lectures on Path Integration: Trieste 1991}, edited by
H.A.Cerdeira {\it et al.} (World Scientific, Singapore, 1993);
L.Schulman and R.W.Ziolkowiski, in {\sl Path integrals from
meV to MeV}, edited by V. Sa-yakanit, W. Sritrakool, J. Berananda, M. C.
Gutzwiller, A. Inomata, S. Lundqvist, J. R. Klauder and L. S. Schulman
(World Scientific, Singapore, 1989);
J.J.Halliwell and M.E.Ortiz, {\sl Phys.Rev.} 
{\bf D48}, 748 (1993).

\refis{Sch1} L.S.Schulman, {\it Time's Arrows and Quantum
Measurement} (Cambridge University Press, Cambridge, 1997).

\refis{Sch2} B.Gaveau and L.S.Schulman, {\sl J.Stat.Phys.}
{\bf 58}, 1209 (1990); L.S.Schulman, {\sl Ann.Phys.} {\bf 212},
315 (1991).

\refis{Time} 
Y.Aharanov and D.Bohm, \pr {\bf 122}, 1649 (1961);
I.Bloch and D.A.Burba, \pr {\bf 10}, 3206 (1974);
V.Delgado, preprint quant-ph/9709037 (1997);
R.Giannitrapani, preprint quant-ph/9611015 (1998);
E.Gurjoy and D.Coon, {\sl Superlattices and 
Microsctructures} {\bf 5}, 305 (1989);
A.S.Holevo, {\it Probabilistic and Statistical Aspects
of Quantum Theory} (North Holland, Amsterdam, 1982), pages 130--197;
D.H.Kobe and V.C.Aguilera--Navarro,  \pr {\bf A50}, 933 (1994);
N.Kumar, {\sl Pramana J.Phys.} {\bf 25}, 363 (1985);
J.Le\'on, preprint quant-ph/9608013 (1996);
L.Mandelstamm and I.Tamm, {\sl J.Phys.} {\bf 9}, 249 (1945);
D.Marolf, \pr {\bf A50}, 939 (1994);
J.G.Muga, J.P.Palao and C.R.Leavens, preprint quant-ph/9803087
(1987);
J.G.Muga, R.Sala and J.P.Palao, preprint quant-ph/9801043,
{\sl Superlattices and Microstructures} {\bf 23} 833 (1998);
C.Piron, in {\it
Interpretation and Foundations of Quantum Theory}, edited by
H.Newmann (Bibliographisches Institute, Mannheim, 1979);
M.Toller, preprint quant-ph/9805030 (1998);
H.Salecker and E.P.Wigner, \pr {\bf 109}, 571 (1958);
F.T.Smith, \pr {\bf 118}, 349 (1960); 
E.P.Wigner, \pr {\bf 98}, 145 (1955).

\refis{YaT} N.Yamada and S.Takagi, {\sl Prog.Theor.Phys.}
{\bf 85}, 985 (1991); {\bf 86}, 599 (1991); {\bf 87}, 77 (1992);
N. Yamada, {\sl Sci. Rep. T\^ohoku Uni., Series 8}, {\bf 12}, 177
(1992); \pr {\bf A54}, 182 (1996).

\endreferences

\end